\documentclass[12pt]{article}
\pdfoutput=1

\usepackage{amsmath}
\usepackage{amsfonts}
\usepackage{amssymb}

\usepackage[
      colorlinks=true,
      linkcolor=blue,
      urlcolor=blue,
      filecolor=black,
      citecolor=red,
      pdfstartview=FitV,
      pdftitle={},
        pdfauthor={ Michael Gutperle, John D. Miller},
        pdfsubject={},
        pdfkeywords={},
        pdfpagemode={},
        bookmarksopen=true
      ]{hyperref}
\hypersetup{linktocpage}

\usepackage{color}
\usepackage{graphicx}

\usepackage{sectsty}
\sectionfont{\large}

\renewcommand{\thefootnote}{\fnsymbol{footnote}}
\renewcommand{\thanks}[1]{\footnote{#1}}
\newcommand{\starttext}{
\setcounter{footnote}{0}
\renewcommand{\thefootnote}{\arabic{footnote}}}
\newcommand{\bea}{\begin{eqnarray}}
\newcommand{\eea}{\end{eqnarray}}
\newcommand{\be}{\begin{equation}}
\newcommand{\ee}{\end{equation}}

\numberwithin{equation}{section}

\usepackage{ dsfont } %for identity matrix

\newcommand{\paren}[1]{\left(#1\right)}
\newcommand{\brak}[1]{\left[#1\right]}

\def\RR{{\mathbb R}}

\long\def\symbolfootnote[#1]#2{\begingroup%
\def\thefootnote{\fnsymbol{footnote}}\footnote[#1]{#2}\endgroup}

\begin{document}
\setlength{\baselineskip}{18pt}

\starttext
\setcounter{footnote}{0}

\begin{flushright}
\today
\end{flushright}

\bigskip

\begin{center}

{\Large \bf Topological interfaces in Chern-Simons theory and $AdS_3/CFT_2$}

\vskip 0.4in

{\large   Michael Gutperle and John D. Miller}

\vskip .2in

{ \it Mani L. Bhaumik Institute for Theoretical Physics }\\
{ \it Department of Physics and Astronomy }\\
{\it University of California, Los Angeles, CA 90095, USA}\\[0.5cm]

\bigskip
\href{mailto:gutperle@physics.ucla.edu}{\texttt{gutperle}}\texttt{, }
\href{mailto:johnmiller@physics.ucla.edu}{\texttt{johnmiller@physics.ucla.edu}}

\bigskip

\bigskip

\end{center}
 
\begin{abstract}

\setlength{\baselineskip}{18pt}

Recently,  topological interfaces between three-dimensional abelian Chern-Simons theories were constructed. In this note we investigate such topological interfaces in the context of the $AdS_3/CFT_2$ correspondence. We show that it is possible to connect the topological interfaces in the bulk Chern-Simons theory to topological interfaces in the dual CFT on the boundary. In addition for  $[U(1)]^{2N}$ Chern-Simons theory  on  $AdS_3$, we show that it is possible  to find boundary counter terms which lead to the $N$ conserved currents in the dual two-dimensional CFT.

\end{abstract}

\setcounter{equation}{0}
\setcounter{footnote}{0}

\newpage

\section{Introduction}

Topological field theories have a wide use in  condensed matter, high energy and mathematical physics.
One of the best-studied examples of a topological field theory is three-dimensional Chern-Simons  (CS) theory \cite{Witten:1988hf}. In the context of the $AdS_3/CFT_2$ correspondence, abelian  CS  theory is entirely responsible for the introduction of objects   in the CFT which are charged under global $U(1)$ currents. The  CS fields  have a natural origin from compactifications of type II or M-theory (see e.g. \cite{Kraus:2006wn}). In the presence of Maxwell kinetic terms the gauge fields decompose  into a massive gauge fields and a flat topological sector \cite{Deser:1981wh}.  Since we are interested in topological questions we do not take the Maxwell terms into account. 
 For the discussion of Maxwell-Chern-Simons theories in the context of AdS/CFT see e.g.  \cite{Gukov:2004id,Chang:2014jna,Hofman:2017vwr}.

When one considers co-dimension one  interfaces between two theories  or boundaries  of a single  theory, the variation of the action can pick up terms localized on the interface or boundary. In order to obtain a good variational principle it may then be necessary to add counter terms  to the action which are 
localized on the interface or boundary. For topological field theories this can lead to the introduction of non-topological degrees of freedom and this procedure  is indeed what causes the relation of CS theory  on a three manifold with boundary and chiral WZW theories on the boundary \cite{Witten:1988hf,ELITZUR1989108}.
On the other hand, as  shown   in \cite{Kapustin:2010hk}, for abelian CS theories it is possible to impose topological boundary conditions, where no counter terms are necessary. Since any interface between two theories can be mapped into a boundary by the folding trick \cite{Bachas:2001vj}  this statement implies the existence of topological interfaces in CS theories \cite{Fliss:2017wop}. The aim of this note is to study some implications of such CS topological interface theories in the context of the AdS/CFT correspondence and relate them to topological interfaces in the dual two-dimensional CFT. 

The structure of this note is as follows: In section  \ref{sec2} we collect background material on CS theories, $AdS_3/CFT_2$ and topological interfaces which will be useful in the main part of the paper.  In section \ref{sec3} we relate a topological interface in the bulk of $AdS_3$ to the  boundary by utilizing an $AdS_2$ slicing of $AdS_3$. In order to identify the conserved  currents in the CFT, they  need to have    both holomorphic and anti-holomorphic parts. To accomplish this we generalize a construction first given in \cite{Keranen:2014ava} to show that in general it is possible to obtain a topological interface in the $CFT_2$ on the boundary from a topological interface in the $AdS_3$ bulk. In section \ref{sec4} we briefly discuss higher dimensional generalizations of this construction. We close with a discussion of open questions in section  \ref{sec5}.

\section{Review of background material}\label{sec2}
In this section we will briefly review well-known material on Chern-Simons (CS) theory, the holographic interpretation of  abelian CS theory in the context $AdS_3/CFT_2$, and topological interfaces in two-dimensional conformal field theories.

\subsection{Chern-Simons theory}

Consider a theory of $N$ abelian gauge fields $A^I, I=1,2,\ldots,N$ on a 3-manifold $\mathcal{M}$, all with period $2\pi$ and with action given by
\begin{equation}\label{pureCSaction}
S_\text{CS}=\frac{K_{IJ}}{4\pi}\int_\mathcal{M}A^I\wedge dA^J
\end{equation}
where $K_{IJ}$ is a symmetric matrix called the level matrix. Following  \cite{Kapustin:2010hk}, we note that the level matrix $K$ has to be integer valued and even for the theory to be well defined on  topologically nontrivial surfaces  under large gauge transformations.  The CS theory is a topological field  theory as the action is independent of a metric on $\mathcal{M}$. The equations of motion following from   (\ref{pureCSaction}) force the connections $A^I$ to be flat 
\begin{equation}
K_{IJ}\,dA^J =0, \quad I=1,2,\ldots,N 
\end{equation}
and hence there are no local propagating degrees of freedom. The only  global gauge invariant observables are Wilson lines. However, for three-dimensional manifolds with boundary there can be nontrivial dynamical fields on the boundary relating three-dimensional CS theory to two-dimensional CFTs \cite{Witten:1988hf}. 
\subsection{Holography for Chern-Simons theory}\label{holocs}
There are several uses for three-dimensional CS theory in $AdS_3/CFT_2$. First, there is the reformulation of three-dimensional gravity in $AdS_3$ in terms of an $SL(2,R)\times SL(2,R)$ CS theory   
  \cite{Witten:1988hc,Achucarro:1987vz} and the subsequent formulation of higher spin gravity as a CS theory (see e.g.  \cite{Bergshoeff:1989ns,Campoleoni:2010zq}). Here we will consider  a different setup, namely the addition of abelian CS matter  to Einstein gravity.
  
Consider an asymptotically $AdS_3$ spacetime in Fefferman-Graham form, with the $AdS_3$ boundary located at $\eta=+\infty$
\begin{equation}
ds^2= d\eta^2 + e^{2\eta \over l} g^{(0)}_{\alpha \beta} dx^\alpha dx^\beta  + g^{(2)}_{\alpha \beta} dx^\alpha dx^\beta + o(e^{-{2\eta \over l}})
\end{equation}
In the gauge $A_\eta^I=0$ the asymptotic form of a gauge field for a general action, including  Maxwell or higher derivative terms, is given by
\begin{equation}
A^I_\alpha = A^I_{(0), \alpha} + e^{-{2\eta \over l}} A^I_{(2), \alpha} + o(e^{-{3\eta \over l}})
\end{equation}
where $A^I_{(0)}$ is flat and only determined through the CS part of the action.  A good variational principle allows us to hold fixed only one boundary component of $ A^I_{(0), \alpha} $. However, the CS action is then not stationary due to the appearance  of a boundary term in the variation. The standard resolution (see e.g. \cite{Kraus:2006wn}) is to add a counter term to the action (\ref{pureCSaction})
\begin{equation}\label{counterterm}
S_{\rm CT} = {1\over 8\pi}K_{IJ}  \int d^2z  \sqrt{-g^{(0)}}\; g^{(0)\;\alpha\beta}  A^I_{(0), \alpha} A^J_{(0), \beta}
\end{equation}
With the addition of this counter term and a flat boundary metric $g^{(0)}_{\alpha\beta}=\eta_{\alpha\beta}$, the variation of the action becomes 
\begin{equation}
\delta S_{\text{total}}=\delta (S_{\rm CS}+S_{\rm CT})= {1\over 2\pi} K_{IJ} \int d^2z  \, A_z^I \delta A_{\bar z}^J
\end{equation}
Hence we can identify $A_{\bar z}$ with the source and the dual current is purely holomorphic
\begin{equation}\label{holocurr}
J_{I,z} = {\delta S_{\text{total}}\over\delta A_{\bar z}^I} ={1\over 2 \pi }\,K_{IJ} A^J_z
\end{equation}
The holomorphic  stress tensor can be obtained from (\ref{counterterm}) and takes the following form 
\begin{equation}\label{holostress}
T_{zz} =  {\pi \over 2}\,K^{IJ} J_{I,z} J_{J,z}
\end{equation}
where $K^{IJ}$ is the inverse of the matrix $K_{IJ}$. If we instead wish to source anti-holomorphic currents, then we instead subtract the counter term (\ref{counterterm}). In this case we can identify $A_z$ with the source, so that the dual current is purely anti-holomorphic
\begin{equation}
J_{I,\bar{z}} = {\delta S_{\rm total}\over\delta A_z^I} =-{1\over 2 \pi }\,K_{IJ} A^J_{\bar{z}}
\end{equation}
and the anti-holomorphic stress tensor takes the form
\begin{equation}
T_{\bar{z}\bar{z}} =  -{\pi \over 2}\,K^{IJ} J_{I,\bar{z}} J_{J,\bar{z}}
\end{equation}

\subsection{Topological interfaces in $CFT_2$}

In two-dimensional CFTs  a  conformal interface is a  one-dimensional line  which separates the two CFTs such that one copy of the conformal symmetry $Virasoro\otimes Virasoro$ is preserved \cite{Bachas:2001vj}. If the interface is localized at $y=0$ in $\RR^2$, a conformal interface satisfies the following continuity condition on the stress tensor components
\begin{equation}\label{confint}
[T^{(L)}_{zz}(x) -T^{(L)}_{\bar z\bar z}(x) ]_{y=0} =  [T^{(R)}_{zz}(x) - T^{(R)}_{\bar z\bar z}(x)]_{y=0}, \quad x\in \RR
\end{equation}
where $T^{(L/R)}$ denotes the stress tensor on the CFT to the left (right)  of the interface, respectively.
There is a special class of conformal interfaces which are called topological, where the continuity holds for the holomorphic and anti-holomorphic components of the stress tensor separately

\begin{equation}\label{topintc}
[T^{(L)}_{zz}(x)  - T^{(R)}_{zz}(x) ]_{y=0}=0,\;\;  \;\; [T^{(L)}_{\bar z\bar z}(x)-  T^{(R)}_{\bar z\bar z}(x) ]_{y=0} =0 , \quad x\in 
\RR
\end{equation}

As argued in \cite{Bachas:2001vj,Petkova:2000ip} a topological  interface can be viewed as an operator which maps $CFT_L$ into $CFT_R$ and the condition (\ref{topintc}) implies that the interface commutes with local conformal transformations and can be continuously deformed.  Topological interfaces are also called totally transmissive interfaces. Topological interfaces have special properties compared to conformal interfaces: a fusion product can be defined when two interfaces come close together \cite{Bachas:2007td}.  Topological interfaces are related to the symmetries of the CFT such as T-duality for a free boson \cite{Bachas:2008jd}. The doubling trick relates an  interface to a boundary in the tensor product $CFT_L\otimes CFT_R$ and for some rational CFTs topological interfaces can be classified using the Cardy construction \cite{Fuchs:2007tx}.

\subsection{Topological interfaces in Chern-Simons theory}
In this section we will review the recent construction of topological interface conditions for CS theory given in \cite{Kapustin:2010hk}\footnote{For related work in the condensed matter literature see, e.g.  \cite{Haldane:1995xgi,Kitaev,Wang:2012am,Levin:2013gaa,Lan:2014uaa}.}. We will be mainly following the treatment given in \cite{Fliss:2017wop}.  We divide  the total 3-manifold $\mathcal{M}$ into two parts  $\mathcal{M}=\mathcal{M}_L\cup_\Sigma\mathcal{M}_R$ with joining interface $\Sigma$. 
The $U(1)^N$ CS action is now divided into two parts 
\begin{equation}\label{leftrighcs}
S_\text{CS}={1\over 4\pi} {K^{(L)}}^{IJ}\int_{\mathcal{M}_L}A^{(L)}_I\wedge dA^{(L)}_J+{1\over 4\pi} {K^{(R)}}^{IJ}\int_{\mathcal{M}_R}A^{(R)}_I\wedge dA^{(R)}_J
\end{equation}
with in general different  level matrices $K^{(L)}$ and $K^{(R)}$. If the manifold $\mathcal{M}$ has a boundary we have to add an appropriate boundary term. In this section we will focus on the topological interface  conditions which relate the $A^{(L)}$ and $A^{(R)}$ gauge fields and postpone the discussion of the boundary terms to section \ref{conscurr}.   

A topological interface is defined such that the canonical symplectic one-form
\begin{equation}
{\bf \Theta}= \delta S_{CS}\mid_{\rm \,on\; shell}\,\,= -{ 1\over 4\pi} \int_{\Sigma } \Big( {K^{(L)}}^{IJ} A^{(L)}_I\wedge \delta A^{(L)}_J- {K^{(R)}}^{IJ} A^{(R)}_I\wedge \delta A^{(R)}_J\Big)
\end{equation}
vanishes on shell on a half-dimensional subspace of the phase space without the introduction of additional contributions coming from counter terms localized on $\Sigma$. These bulk boundary conditions are determined by two $N\times N$ matrices $v^{(L)}$ and $v^{(R)}$ which implement the boundary condition
\begin{equation}\label{bulkTBCs}
{v^{(L)}}^\text{T}K^{(L)}A^{(L)}\big|_\Sigma=-{v^{(R)}}^\text{T}K^{(R)}A^{(R)}\big|_\Sigma
\end{equation}
and must respect the gluing condition
\begin{equation}\label{gluecond}
{v^{(L)}}^\text{T}K^{(L)}v^{(L)}={v^{(R)}}^\text{T}K^{(R)} v^{(R)}
\end{equation}
Since the above gluing condition does not have unique solutions, we additionally demand that the $v^{(L)}$ and $v^{(R)}$ satisfy a primitivity condition. This translates to the condition that the $N\times N$ minors of the $2N\times N$ matrix
\begin{equation}
\mathbb{P}=\begin{pmatrix}
v^{(L)} \\ -v^{(R)}
\end{pmatrix}
\end{equation}
all have a greatest common divisor of 1.

As an example of interface conditions between theories with unequal level matrices, consider the case
\begin{equation}\label{stndKmtx}
K^{(L)}=kn_L^2\begin{pmatrix}
1 & 0 \\ 0 & -1 \end{pmatrix}\,\,\,\,\,\,\text{and}\,\,\,\,\,\,K^{(R)}=kn_R^2\begin{pmatrix}
1 & 0 \\ 0 & -1 \end{pmatrix}
\end{equation}
where $k,n_L,n_R\in\mathbb{Z}$ and we assume that $n_L$ and $n_R$ are relatively prime. There are two types of primitive boundary condition matrices satisfying (\ref{gluecond}), either
\begin{equation}
v_1^{(L)}=n_R\begin{pmatrix} \eta_L & 0 \\ 0 & \eta'_L \end{pmatrix}\,\,\,\,\,\,\text{and}\,\,\,\,\,\,v_1^{(R)}=n_L\begin{pmatrix} \eta_R & 0 \\ 0 & \eta'_R \end{pmatrix}
\end{equation}
or
\begin{equation}
v_2^{(L)}=n_R\begin{pmatrix} 0 & \eta''_L \\ \eta''_L & 0 \end{pmatrix}\,\,\,\,\,\,\text{and}\,\,\,\,\,\,v_2^{(R)}=n_L\begin{pmatrix} 0 & \eta''_R \\ \eta''_R & 0 \end{pmatrix}
\end{equation}
where $\eta_L,\eta'_L,\eta''_L,\eta_R,\eta'_R,\eta''_R=\pm 1$. In terms of the boundary condition
\begin{equation}
A^{(L)}\big|_\Sigma=-v^{(L)}[v^{(R)}]^{-1}A^{(R)}\big|_\Sigma
\end{equation}
following from (\ref{bulkTBCs}) and (\ref{gluecond}), we have that
\begin{equation}\label{stndgBCs}
-v_1^{(L)}[v_1^{(R)}]^{-1}=-\frac{n_R}{n_L}\begin{pmatrix}
\eta_L\eta_R & 0 \\ 0 & \eta'_L\eta'_R
\end{pmatrix}\,\,\,\,\,\,\text{and}\,\,\,\,\,\,-v_2^{(L)}[v_2^{(R)}]^{-1}=-\frac{n_R}{n_L}\begin{pmatrix}
\eta''_L\eta''_R & 0 \\ 0 & \eta''_L\eta''_R
\end{pmatrix}
\end{equation}
While the diagonal level matrices of (\ref{stndKmtx}) do not allow for boundary conditions that mix the gauge fields of different levels, in general diagonal level matrices do. For example \cite{Fliss:2017wop}, the continuous level matrix
\begin{equation}\label{ueKmtx}
K^{(L)}=K^{(R)}=k\begin{pmatrix} 1 & 0 \\ 0 & m^2-n^2 \end{pmatrix}
\end{equation}
with $n,m$ relatively prime, allows for the primitive boundary condition matrices
\begin{equation}\label{uegBCs}
v^{(L)}=\begin{pmatrix} m & n \\ 0 & 1 \end{pmatrix}\,\,\,\,\,\,\text{and}\,\,\,\,\,\,v^{(R)}=\begin{pmatrix} -n & -m \\ 1 & 0 \end{pmatrix}\,\,\Longrightarrow\,\,-v^{(L)}[v^{(R)}]^{-1}=\frac{1}{m}\begin{pmatrix} n & n^2-m^2 \\ 1 & n \end{pmatrix}
\end{equation}

\section{Topological interfaces in the $AdS$ bulk}
\label{sec3}
In this section of the note we discuss how  a topological interface in the bulk CS theory can be related to topological interfaces in two-dimensional CFTs via the AdS/CFT correspondence. 

\subsection{$AdS_2$ slicing}

\begin{figure}[t]
\centering
\includegraphics[scale=1.0]{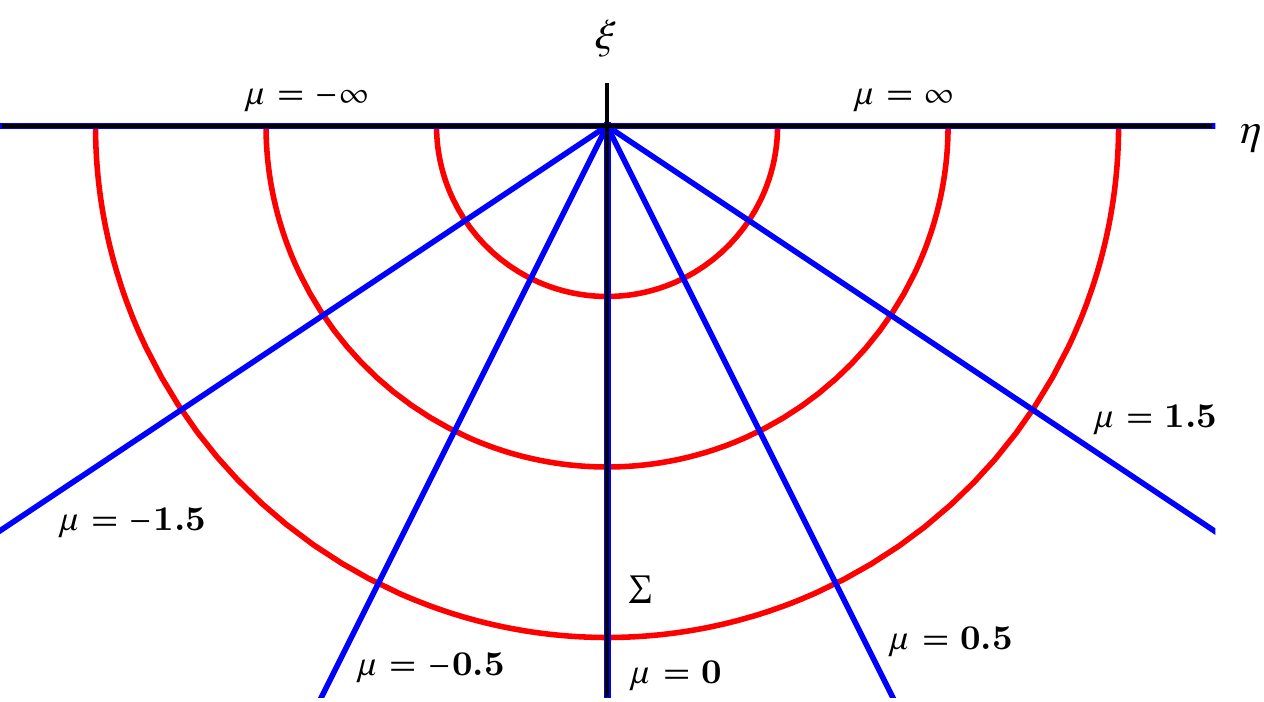}
\caption{$AdS_2$ slicing in Poincare coordinates. Curves of constant $\mu$ (blue) and $x$ (red) illustrate how the spatial boundary coordinate on one side of the boundary interface is extends through the bulk to the opposite side of the boundary interface.}\label{ads2slicefig}
\end{figure}

A useful coordinate system to work with is that of an $AdS_2$ slicing of $AdS_3$, which has been used in the construction of Janus solutions before \cite{Bak:2003jk}
\begin{equation}\label{adsslicing}
ds^2= d\mu^2 + \cosh^2 {\mu } \,{dx^2-dt^2\over x^2} 
\end{equation}
The boundary of $AdS_3$ consists of three components: two half-spaces reached by taking $\mu \to \pm \infty$ and the boundary of $AdS_2$ reached by taking $x\to 0$. While it seems that the three conformal boundary components are disconnected this is an artifact of the coordinate system which can be seen by mapping the metric (\ref{adsslicing}) to the standard Poincare slicing $AdS_3$ metric
\begin{equation}\label{Poncslicing}
ds^2= {1\over \xi^2} \left(d\xi^2+ d\eta^2-dt^2\right)
\end{equation}
via the coordinate transformation
\begin{equation}
\mu= \tanh^{-1}\left({\eta\over \sqrt{\xi^2+\eta^2}}\right), \quad x= \sqrt{\xi^2+\eta^2}
\end{equation}
which shows that the boundary half-spaces  $\mu \to \pm \infty$ are glued together at the interface $x=0$.  In the 
coordinate system (\ref{adsslicing})  we locate the CS topological interface at $\mu=0$ and $
\mathcal{M}_{L/R}$ are given by the half-spaces $\mu<0$ and $\mu>0$ respectively (see figure \ref{ads2slicefig}). In this coordinate system we 
can impose the gauge $A_\mu^I=0$. It then follows from the flatness of the connection that the non-vanishing components $A_{z,t}$ are independent of $\mu$ and hence the connection at the CS 
interface can be trivially related to the connection at the boundary component of $AdS_3$. Note  that 
due to the fact that the CS action is topological there is no backreaction on the metric, which remains 
unchanged from (\ref{Poncslicing}). This is to be contrasted to the case of Janus solutions involving 
massless scalars \cite{Bak:2003jk,DHoker:2007zhm}, where the metric will be deformed.

\subsection{Simple holomorphic example}

In the following we consider the  boundary counter terms  discussed in section \ref{holocs}, which lead to  purely holomorphic $U(1)$ currents (\ref{holocurr}) and stress tensor (\ref{holostress}).
We utilize the $AdS_2$ slicing coordinates given in (\ref{adsslicing}) and locate the topological interface in the bulk at $\mu=0$ with the  left and right CS theories in (\ref{leftrighcs}) occupying $\mu<0$ and $\mu>0$ respectively. As discussed in the last section,  the gauge $A_\mu=0$ allows for $A_{a}, a=x,t$,  to  be trivially continued to the $AdS_3$ boundary at $\mu=\pm \infty$ and compared at the location of the $CFT_2$ interface at $x=0$.  Using this, the matching condition (\ref{bulkTBCs}) at the bulk topological interface translates into the following condition for the currents
\be
\left( v^{(L)} \right )^T J^{(L)}_z \mid_{x=0} \,\,= - \left( v^{(R)} \right )^T J^{(R)}_z \mid_{x=0} 
\ee
We can use this matching condition to relate the holomorphic stress tensor for the left and right CFTs
\bea
\left( J^{(L)}\right)^T K_{(L)}^{-1} J^{(L)} 
&=&  \left(J^{(R)}_z \right)^T     v^{(R)}  \left( v^{(L)}\right )^{-1} K_{(L)}^{-1}  \left( (v^{(L)})^T\right )^{-1}  \left( v^{(R)} \right )^T J^{(R)}_z \nonumber\\
&=&  \left(J^{(R)}_z \right)^T   v^{(R)}   \left(  (v^{(L)})^T K_{(L)} v^{(L)}\right)^{-1} \left( v^{(R)} \right )^T J^{(R)}_z\nonumber\\
&=&   \left(J^{(R)}_z \right)^T   v^{(R)}   \left(  (v^{(R)})^T K_{(R)} v^{(R)}\right)^{-1} \left( v^{(R)} \right )^T J^{(R)}_z\nonumber\\
&=&  \left(J^{(R)}_z \right)^T   K_{(R)}^{-1}  J^{(R)}_z
\eea
where in the last line we used the gluing condition (\ref{gluecond}) for the $K$ matrices.  It follows from the definition (\ref{holostress}) that the holomorphic components of the stress tensor are continuous
\be\label{holconda}
T_{zz}^L\mid_{x=0} = T_{zz}^R\mid_{x=0}
\ee
which is the first condition in (\ref{topintc}) a topological CFT  interface must satisfy. However, in the purely holomorphic formulation discussed so far it is not possible to construct the anti-holomorphic stress tensor and hence verify the second condition in (\ref{topintc}).\footnote{If the interface condition  is conformal and satisfies (\ref{confint}) the holomorphic condition (\ref{holconda}) implies the anti-holomorphic one.} Even when the level matrices decompose according to $K^{(L,R)}=k^{(L,R)}\oplus(-\tilde{k}^{(L,R)})$, where we can choose to source holomorphic currents from the gauge fields mixed by $k^{(L,R)}$ and source anti-holomorphic currents from the gauge fields mixed by $-\tilde{k}^{(L,R)}$, there are problems with the continuity of the stress tensor components. To see this, let us write
\begin{equation}
A^{(L,R)}\big|_{\partial\mathcal{M}_{L,R}}=\begin{pmatrix}
a^{(L,R)} \\ \tilde{a}^{(L,R)}
\end{pmatrix}
\end{equation}
so that we have
\begin{equation}
J^{(L,R)}_z=\frac{1}{2\pi}\,k^{(L,R)}a^{(L,R)}_z\,\,\,\,\,\,\text{and}\,\,\,\,\,\,\tilde{J}^{(L,R)}_{\bar{z}}=\frac{1}{2\pi}\,\tilde{k}^{(L,R)}\tilde{a}^{(L,R)}_{\bar{z}}
\end{equation}
and the stress tensor components are given by
\begin{equation}\label{naiveholostress}
T^{(L,R)}_{zz}=\frac{\pi}{2}\,J_z^{(L,R)}(k^{(L,R)})^{-1}J_z^{(L,R)}+\frac{1}{8\pi}\,\tilde{a}_z^{(L,R)}\tilde{k}^{(L,R)}\tilde{a}_z^{(L,R)}
\end{equation}
\begin{equation}\label{naiveaholostress}
T^{(L,R)}_{\bar{z}\bar{z}}=\frac{1}{8\pi}\,a_{\bar{z}}^{(L,R)}k^{(L,R)}a_{\bar{z}}^{(L,R)}+\frac{\pi}{2}\,\tilde{J}_{\bar{z}}^{(L,R)}(\tilde{k}^{(L,R)})^{-1}\tilde{J}_{\bar{z}}^{(L,R)}
\end{equation}
One can check that (\ref{naiveholostress}) and (\ref{naiveaholostress}) are separately continuous for the boundary conditions (\ref{stndgBCs}), but not for those of (\ref{uegBCs}). Generally, the stress tensor components produced by these counter terms will only be separately continuous if the boundary conditions decompose according to
\begin{equation}
-v^{(L)}{v^{(R)}}^{-1}=\begin{pmatrix} V & 0 \\ 0 & \tilde{V} \end{pmatrix}
\end{equation}
which from the boundary conditions on $a^{(L,R)}$ on $\partial\Sigma$
\begin{equation}
{v^{(L)}}^\text{T}\begin{pmatrix} 2\pi J^{(L)}_z\\ -\tilde{k}^{(L)}\tilde{a}^{(L)}_z \end{pmatrix}\Big|_{\partial\Sigma}=-{v^{(R)}}^\text{T}\begin{pmatrix} 2\pi J^{(R)}_z\\ -\tilde{k}^{(R)}\tilde{a}^{(R)}_z \end{pmatrix}\Big|_{\partial\Sigma}
\end{equation}
\begin{equation}
{v^{(L)}}^\text{T}\begin{pmatrix} k^{(L)}a^{(L)}_{\bar{z}} \\ -2\pi \tilde{J}^{(L)}_{\bar{z}} \end{pmatrix}\Big|_{\partial\Sigma}=-{v^{(R)}}^\text{T}\begin{pmatrix} k^{(R)}a^{(R)}_{\bar{z}} \\ -2\pi \tilde{J}^{(R)}_{\bar{z}} \end{pmatrix}\Big|_{\partial\Sigma}
\end{equation}
we see is related to the possible mixing between holomorphic and anti-holomorphic boundary currents and the remaining components of the bulk fields. With counter term choices like (\ref{counterterm}) we will always have this problem owing to the fact that the holomorphic and anti-holomorphic currents are independent from each other. This is the reason why we generalize the counter terms in the next section in order to obtain a conserved current with both holomorphic and anti-holomorphic parts.

\subsection{Pure CS counter terms and conserved currents}\label{conscurr}

In \cite{Keranen:2014ava}, an interesting counter term was chosen in order for the bulk CS theory to be sourced by a boundary current whose components were not separately conserved. Such a boundary current then has no chiral anomaly as a result of the flatness of the gauge fields which it sources. Specifically, the action of the theory is given by
\begin{equation}\label{KeranenCT}
S=\frac{k}{4\pi}\int_\mathcal{M}\paren{A\wedge dA-\bar{A}\wedge d\bar{A}}+\frac{k}{8\pi}\int_{\partial\mathcal{M}} d^2z\paren{A_zA_{\bar{z}}+\bar{A}_z\bar{A}_{\bar{z}}-2A_{\bar{z}}\bar{A}_z}
\end{equation}
where the first two terms in the counter term allow $A_{\bar{z}}$ and $\bar{A}_z$ to be fixed on the boundary and the final term is chosen to produce a conserved current; i.e. the boundary currents
\begin{equation}
J_z=\frac{\delta S}{\delta A_{\bar{z}}}=\frac{k}{2\pi}\paren{A_z-\bar{A}_z}\Big|_{\partial\mathcal{M}}\,\,\,\,\,\,\text{and}\,\,\,\,\,\,J_{\bar{z}}=\frac{\delta S}{\delta \bar{A}_z}=-\frac{k}{2\pi}\paren{A_{\bar{z}}-\bar{A}_{\bar{z}}}\Big|_{\partial\mathcal{M}}
\end{equation}
can be regarded as components of a single current satisfying
\begin{equation}
\partial^\mu J_\mu=\frac{k}{2\pi}\brak{\paren{\partial_{\bar{z}}A_z-\partial_zA_{\bar{z}}}-\paren{\partial_{\bar{z}}\bar{A}_z-\partial_z\bar{A}_{\bar{z}}}}\Big|_{\partial\mathcal{M}}=0
\end{equation}
by the flatness of $A$ and $\bar{A}$. If we want $A$ and $\bar{A}$ to be sourced by left- and right- moving currents, respectively, then (\ref{KeranenCT}) is the unique counter term for which such a conserved current can be constructed; however, if we make no assumptions about which gauge fields source the left-moving and right-moving currents then larger classes of counter terms are possible.

Consider the pure CS action (\ref{pureCSaction}) of $2N$ gauge fields in $AdS_3$, with the addition of a generic quadratic counter term. Making use of the Hodge star on $\partial\mathcal{M}$, we can write such a counter term in the coordinate invariant form
\begin{equation}\label{genCTform}
S_\text{CT}=\frac{1}{8\pi}\int_{\partial\mathcal{M}}\paren{X_{IJ}*A^I\wedge A^J+Y_{IJ}\,A^I\wedge A^J}
\end{equation}
where here $X$ and $Y$ are symmetric and anti-symmetric $2N\times 2N$ matrices, respectively. The variation of the total action is then given by
\begin{equation}\label{totalvarform}
\delta S_\text{total}=\frac{1}{4\pi}\int_{\partial\mathcal{M}}*\brak{X_{IJ}A^I+\paren{K_{IJ}+Y_{IJ}}*A^I}\wedge\delta A^J
\end{equation}
Decomposing the term in the brackets above in terms of its self-dual and anti-self-dual parts, we see that in order to allow for a well-defined variational principle consistent with $N$ left-moving and $N$ right-moving boundary currents it must be the case that the matrices
\begin{equation}\label{Pdef}
P_\pm=X\pm Y\pm K
\end{equation}
each be half-rank. Furthermore, the nullspaces of these matrices and their transposes determine the boundary currents and the gauge fields they source. Specifically, the left- and right- moving boundary currents will be combinations of the gauge fields valued in the orthogonal complements to the nullspaces $N_\pm^\text{T}$ of $P_\pm^\text{T}$; and the combinations of the gauge fields sourcing them will be valued in the orthogonal complements to the nullspaces $N_\pm$ of $P_\pm$. Thus, for a well-defined variational principle we must specifically have that $N_++N_-=\mathbb{R}^{2N}$, and for it to be possible to construct $N$ conserved currents we must have $N_+^\text{T}=N_-^\text{T}$. Such matrices can be constructed from a spanning set of vectors $\{v^+_i,v^-_i\}$ and another set of linearly independent vectors $\{w_i\}$ and setting
\begin{equation}\label{Pconstruct1}
P_\pm^T=\sum_{i=1}^N v^\pm_iw_i^\text{T}
\end{equation}
where the $\{v^\pm_i\}$ form bases for the orthogonal complements to $N_\pm$ and the $\{w_i\}$ form a basis for the orthogonal complement to $N_\pm^\text{T}$. Furthermore, consistency with (\ref{Pdef}) constrains the possible vectors in (\ref{Pconstruct1}). First, if
\begin{equation}
X=\frac{1}{2}\paren{P^\text{T}_++P^\text{T}_-}
\end{equation}
is to be a symmetric matrix then we must set $w_i=v^+_i+v^-_i$. Then, writing $K$ in spectral form as
\begin{equation}
K=\sum_{i=1}^N\paren{k^+_iu^+_i{u^+_i}^\text{T}-k^-_iu^-_i{u^-_i}^\text{T}}
\end{equation}
where $u^\pm_i$ are the unit eigenvectors corresponding to the positive and negative eigenvalues $\pm k^\pm_i$ of $K$, we see that
\begin{equation}
K=\frac{1}{2}\brak{\frac{1}{2}\paren{P^\text{T}_+-P^\text{T}_-}+\frac{1}{2}\paren{P_+-P_-}}=\frac{1}{2}\sum_{i=1}^N\paren{v^+_i{v^+_i}^\text{T}-v^-_i{v^-_i}^\text{T}}
\end{equation}
determines the possible $\{v^\pm_i\}$ to be given by
\begin{equation}\label{CTsolution}
\begin{pmatrix} {v^+_i}^\text{T} \\ {v^-_i}^\text{T} \end{pmatrix}=M\begin{pmatrix} \sqrt{2k_i^+}\,{u^+_i}^\text{T} \\ \sqrt{2k_i^-}\,{u^-_i}^\text{T} \end{pmatrix}
\end{equation}
where $M$ is an arbitrary $O(N,N)$ matrix acting on the $\{I\}$ coordinates in some ordering $\{i,i+N\}$.

In terms of the solution (\ref{CTsolution}), the variation (\ref{totalvarform}) can be written as
\begin{equation}
\delta S=\int_{\partial\mathcal{M}}\brak{*\paren{J^i+*J^i}\wedge\delta A_i+*\paren{J^i-*J^i}\wedge\delta\bar{A}_i}
\end{equation}
with the fields sourcing the self-dual and anti-self-dual currents being
\begin{equation}\label{AAbdef}
A_i=c_i\paren{v^+_i}_IA^I=c_i\paren{MUA}_i\,\,\,\,\,\,\text{and}\,\,\,\,\,\,\bar{A}_i=-c_i\paren{v^-_i}_IA^I=-c_i\paren{MUA}_{i+N}
\end{equation}
where the $c_i$ are arbitrary proportionality constants and the matrix $U$ is constructed row-wise as
\begin{equation}
U=\begin{pmatrix} \sqrt{2k_i^+}\,{u^+_i}^\text{T} \\ \sqrt{2k_i^-}\,{u^-_i}^\text{T} \end{pmatrix}
\end{equation}
In terms of the $A_i$ and $\bar{A}_i$, the currents are given by
\begin{equation}\label{Jdef}
J_i=*\,\frac{1}{2\pi c_i^2}\paren{A_i-\bar{A}_i}\Big|_{\partial\mathcal{M}}
\end{equation}
As advertised, we have that
\begin{equation}
d*J_i=\frac{1}{2\pi c_i^2}\paren{dA_i-d\bar{A}_i}\Big|_{\partial\mathcal{M}}=0
\end{equation}
by the flatness of the gauge fields. In terms of (\ref{AAbdef}) and (\ref{Jdef}), the counter term can be written as
\begin{equation}
S_\text{CT}=\int_{\partial\mathcal{M}}\brak{\frac{\pi c_i^2}{2}\,J^i\wedge *\,J_i+\frac{1}{4\pi c_i^2}\,A^i\wedge\bar{A}_i}
\end{equation}
from which we see that the stress tensor is given by
\begin{equation}
T_{\mu\nu}=-\frac{\pi c_i^2}{2}\paren{J^i_\mu J^{\,}_{i,\nu}-\tfrac{1}{2}\,g_{\mu\nu}J^i_\lambda J^{\lambda}_i}
\end{equation}
In flat coordinates, the non-zero components are
\begin{equation}\label{genstress}
T_{zz}=-\frac{\pi c_i^2}{2}\,J^i_zJ^{\,}_{i,z}\,\,\,\,\,\,\text{and}\,\,\,\,\,\,T_{\bar{z}\bar{z}}=-\frac{\pi c_i^2}{2}\,J^i_{\bar{z}}J^{\,}_{i,\bar{z}}
\end{equation}

\subsection{Interfaces with conserved currents}

In order for an interface to preserve the stress tensor components (\ref{genstress}), the boundary conditions on the gauge fields must act as an $O(N)$ transformation on the $c^iJ^i$. Specifically, if the boundary conditions on the fields are
\begin{equation}\label{AIbc}
A_I^{(L)}=\Lambda^J_IA_J^{(R)}
\end{equation}
then we are concerned with the matrix $\hat{\Lambda}$ implementing the conditions on the $c_i^{-1}A_i$ and $c_i^{-1}\bar{A}_i$,
\begin{equation}
\begin{pmatrix}
\tfrac{1}{c_i^{(L)}}\big(A_i^{(L)}-\bar{A}_i^{(L)}\big) \\ \tfrac{1}{c_i^{(L)}}\big(A_i^{(L)}+\bar{A}_i^{(L)}\big)
\end{pmatrix}=\hat{\Lambda}\begin{pmatrix}
\tfrac{1}{c_i^{(R)}}\big(A_i^{(R)}-\bar{A}_i^{(R)}\big) \\ \tfrac{1}{c_i^{(R)}}\big(A_i^{(R)}+\bar{A}_i^{(R)}\big)
\end{pmatrix}
\end{equation}
given by
\begin{equation}
\hat{\Lambda}=\frac{1}{2}\begin{pmatrix} 1 & 1 \\ 1 & -1 \end{pmatrix}M^{(L)}U^{(L)}\Lambda\big(M^{(R)}U^{(R)}\big)^{-1}\begin{pmatrix} 1 & 1 \\ 1 & -1 \end{pmatrix}
\end{equation}
Thus, in order for the boundary conditions (\ref{AIbc}) to act as
\begin{equation}
c_i^{(L)}J_i^{(L)}=S^j_i\,c_j^{(R)}J_j^{(R)}
\end{equation}
for some $S\in O(N)$, the matrix $\hat{\Lambda}$ must decompose according to
\begin{equation}\label{extratopconds}
\hat{\Lambda}=\begin{pmatrix}
S & 0 \\ \hat{\Lambda}_{21} & \hat{\Lambda}_{22}
\end{pmatrix}
\end{equation}
Writing (\ref{gluecond}) in terms of $\Lambda$ and utilizing the spectral decomposition of the level matrices, we see that the combination
\begin{equation}
M_\Lambda=U^{(L)}\Lambda\big(U^{(R)}\big)^{-1}
\end{equation}
is always an $O(N,N)$ matrix, from which fact we determine that all solutions obeying (\ref{extratopconds}) are given by
\begin{equation}\label{CTmatching}
M^{(L)}M_\Lambda\big(M^{(R)}\big)^{-1}=\begin{pmatrix}
S & 0 \\ 0 & S
\end{pmatrix}
\end{equation}
The above shows that there is always enough freedom in the choice of counter terms on the left and right theories to produce a continuous boundary stress tensor.

As an example, for $N=1$ (\ref{CTmatching}) implies that
\begin{equation}\label{CTmatching1}
M_\Lambda=\pm M^{\eta^{(L)}_+\eta^{(R)}_+}_{\eta^{(L)}_-\eta^{(R)}_-}(\lambda^{(R)}-\lambda^{(L)})
\end{equation}
where
\begin{equation}
M^{\eta_+}_{\eta_-}(\lambda)=\begin{pmatrix}
\cosh\lambda & \sinh\lambda \\ \sinh\lambda & \cosh\lambda 
\end{pmatrix}\begin{pmatrix}
\eta_+ & 0 \\ 0 & \eta_-
\end{pmatrix}
\end{equation}
is a general $O(1,1)$ element. We will consider two examples of $N=1$ bulk interfaces, the first of which are
\begin{equation}
M_\Lambda=\begin{pmatrix}
\eta_1 & 0 \\ 0 & \eta_2
\end{pmatrix}
\end{equation}
for the boundary conditions respecting the gluing conditions of the level matrices (\ref{stndKmtx}), where $\eta_1,\eta_2=\pm 1$. In order for (\ref{CTmatching1}) to be obeyed, we must have $\eta^{(L)}_+\eta^{(R)}_+=\pm\eta_1$, $\eta^{(L)}_-\eta^{(R)}_-=\pm\eta_2$, and $\lambda^{(L)}=\lambda^{(R)}$. As a second example, we consider
\begin{equation}
M_\Lambda=\begin{pmatrix}
\tfrac{n}{m} & \sqrt{\tfrac{n^2}{m^2}-1} \\ \sqrt{\tfrac{n^2}{m^2}-1} & \tfrac{n}{m}
\end{pmatrix}
\end{equation}
for the boundary conditions respecting the level matrices (\ref{ueKmtx}). This time, the condition (\ref{CTmatching1}) sets $\eta^{(L)}_+=\eta^{(L)}_-=\pm\eta^{(R)}_+=\pm\eta^{(R)}_-$ and $\lambda^{(R)}-\lambda^{(L)}=\text{arccosh}\,(n/m)$.

\section{Higher-dimensional generalizations}\label{sec4}

We can consider higher-dimensional generalizations of three dimensional CS topological field theory. The most straight forward generalization exists in  $d=4n+3$ dimensions with $n\geq 1$, utilizing $(2n+1)$-dimensional antisymmetric tensor fields
\be
S={K_{IJ} \over 4\pi } \int_{{\cal M}_{4n+3}} B^I \wedge dB^J
\ee
For $n=1$  the matrix $K$ is symmetric just as for the three-dimensional  CS theory, and the theory describes the topological sector of $(2,0)$ theories on M5-branes. This topological field theory has been studied in the past, see e.g.   \cite{Verlinde:1995mz,Belov:2006jd,Heckman:2017uxe}.  Following the 3d example we can consider an $(4n+2)$-dimensional interface $\Sigma$  separating two AST theories with different $K$ matrices living on ${\cal M}_{L,R}$ respectively\footnote{For theories in $d=4n+1$ with $2n$-dimensional AST fields, the matrix $K$ is anti-symmetric and the analysis of topological interface theories does not parallel the CS case.}
\be\label{Sintseven}
S_{\rm int}={K^{(L)}_{IJ} \over 4\pi } \int_{{\cal M}_L}  B^I_{(L)} \wedge dB^J_{(L)}+ {K^{(R)} _{IJ} \over 4\pi } \int_{{\cal M}_R}  B^I_{(R)} \wedge dB^J_{(R)}
\ee
A topological interface with a good variational principle would, as before, have a vanishing symplectic one-form 
\begin{equation}
{\bf \Theta}= \delta S \mid_{\rm \,on\; shell}\,\,= -{ 1\over 4\pi} \int_{\Sigma } \Big( {K^{(L)}}^{IJ} B^{(L)}_I\wedge \delta B^{(L)}_J- {K^{(R)}}^{IJ} B^{(R)}_I\wedge \delta B^{(R)}_J\Big)+ \bf \Theta_{\rm CT}
\end{equation}
with matching conditions which restrict the AST fields  to a half-dimensional Lagrangian subspace. 
A topological interface condition is given when no counter terms which depend on the induced metric on the interface $\Sigma$ have to be added.   While there are many mathematical subtleties in the exact treatment of these theories \cite{Belov:2006jd,Freed:2012bs} it seems likely that topological interfaces can be constructed for these theories, and it would be interesting to investigate what would be the analog of the two-dimensional topological interfaces for the boundary theories when (\ref{Sintseven}) is placed in $AdS_{4n+3}$. 

\section{Discussion}\label{sec5}

In this brief note  we placed abelian three-dimensional CS theories in $AdS_3$ and related the  topological 
interfaces in this theory to topological interfaces in the boundary CFT.  In order to obtain both holomorphic and anti-holomorphic currents and stress tensors, we generalized a construction which produces conserved $U(1)$ currents with both holomorphic and anti-holomorphic  components in the boundary. There are many open questions which would be interesting to pursue.
The relation between CS theories and rational CFTs generalizes to non-abelian CS theories (and WZW models); does the relation of topological interfaces in bulk and boundary theories generalize to this case? The first step in answering this question involves generalizing the classification of topological interfaces in abelian CS theories \cite{Kapustin:2010hk} to the non-abelian case.
One very important property of topological interfaces in two-dimensional CFTs is that they have a nontrivial fusion product, which can be constructed by bringing two topological interfaces close together. 
It would be interesting to understand what the analog of this product is on the bulk side.
The higher-dimensional generalization is also very interesting, in particular whether the topological interfaces -- if they can be consistently defined -- have any interpretation or application in the M5-brane $(2,0)$ theory. We leave the investigation of these questions for future work.

\section*{Acknowledgements}
We are grateful to Per Kraus for useful conversations.
The work of M.G.  is supported in part by the National Science Foundation under grant PHY-16-19926. J.D.M. is grateful to the Mani L. Bhaumik Institute for Theoretical Physics for support.

\end{document}